\documentclass[a4paper]{jpconf}

\usepackage{amssymb}
\usepackage{amsmath}
\usepackage{booktabs}

\usepackage{graphicx}
\graphicspath{ {./Plots/} }

\newcommand*\ops{{\ensuremath{\text{o-Ps}}}}
\newcommand*\pps{{\ensuremath{\text{p-Ps}}}}

\usepackage[detect-all]{siunitx}
	\DeclareSIUnit\gauss{G}											
	\DeclareSIUnit\positron{\text{\ensuremath{\text{e}^+}}}			
	\DeclareSIUnit\electron{\text{\ensuremath{\text{e}^-}}}			
	\DeclareSIUnit\ppm{\text{ppm}}									
	\DeclareSIUnit\ppb{\text{ppb}}									
\usepackage{physics}

\usepackage[caption=false]{subfig}

\usepackage{multirow}

\usepackage{hyperref}
\hypersetup
{
	pdfauthor = {Carlos Vigo},
	pdftitle = {A new approach for the Ps lifetime determination in a vacuum cavity},
}

\begin{document}

\title{A new approach for the ortho-positronium lifetime determination in a vacuum cavity}

\author{C.~Vigo, M.~Raajimakers, L. Gerchow, B.~Radics, A.~Rubbia and P.~Crivelli}
\address{ETH Zurich, Institute for Particle Physics and Astrophysics}
\ead{carlosv@ethz.ch, paolo.crivelli@cern.ch}

\begin{abstract}
Currently, the experimental uncertainty for the determination of the ortho-positronium (\ops\ ) decay rate is at \SI{150}{\ppm} precision; this is two orders of magnitude lower than the theoretical one, at \SI{1}{\ppm} level. Here we propose a new proof of concept experiment aiming for an accuracy of \SI{100}{\ppm} to be able to test the second-order correction in the calculations, which is $\simeq 45\left(\frac{\alpha}{\pi}\right)^2\approx\SI{200}{\ppm}$. The improvement relies on a new technique to confine the \ops\ in a vacuum cavity. Moreover, a new method was developed to subtract the time dependent pick-off annihilation rate of the fast backscattered positronium from the \ops\ decay rate prior to fitting the distribution. Therefore, this measurement will be free from the systematic errors present in the previous experiments. The same experimental setup developed for our recent search for invisible decay of ortho-positronium is being used. The precision will be limited by the statistical uncertainty, thus, if the expectations are fulfilled, this experiment could pave the way to reach the ultimate accuracy of a few \si{ppm} level to confirm or confront directly the higher order QED corrections.  This will provide a sensitive test for new physics, e.g. a discrepancy between theoretical prediction and measurements could hint the existence of an hidden sector which is a possible dark matter candidate.
\end{abstract}

\section{\label{sec:Intro}Introduction}

The bound state between an electron and a positron, positronium (Ps), is a great tool to test the predictions of quantum electrodynamics (QED). Its properties can be perturbatively calculated to high precision using bound state QED. Unlike the hydrogen atom, Ps being a purely leptonic system, is not affected by finite size or QCD effects at current experimental precision level. A review of bound state QED of simple atoms can be found in Ref.~\cite{Karshenboim2005}. The theoretical value of the decay rate of the triplet state, ortho-positronium (\ops), has recently been improved by calculating the higher order $\mathcal{O}\left(\alpha^2\right)$ and logarithmic correction terms to yield $\lambda_\text{T} = \SI{7.039970(10)}{\per\micro\second}$ (\SI{1.5}{\ppm})~\cite{Adkins2000, Adkins2002, Kniehl2000, Hill2000, Melnikov2000}, with research ongoing~\cite{Adkins2015}.

The measured decay rate had a long history of inconsistency with the theoretical predictions, the so-called \ops\ lifetime puzzle. The latest experiments of the decay rate are in agreement with the theoretical predictions~\cite{Vallery2003, Asai1995, Kataoka2009}. Measuring the decay rate to below 100 ppm precision would directly test the $\mathcal{O}\left(\alpha^2\right)$ (\SI{246}{\ppm}) correction of QED. The logarithmic terms are at \SI{4}{\ppm} and beyond the scope of this study. A discrepancy compared to the theoretical prediction could indicate new physics, e.g.~a hidden sector which is a possible dark matter candidate~\cite{Crivelli2010, Vigo2018}.

One of the main challenges in the \ops\ lifetime measurement is that a fraction of the produced atoms will annihilate via pick-off annihilations due to collisions with electrons of the target material or the vacuum cavity walls. Therefore, the observed \ops\ decay rate $\lambda_\text{obs}$ is a sum of the intrinsic \ops\ decay rate $\lambda_{\ops}$ and the pick-off annihilation rate into $2\gamma$'s, $\lambda_\text{pick}$:

\begin{equation}
	\lambda_\text{obs}\left(t\right)=\lambda_{\ops}+\lambda_\text{pick}\left(t\right)
\end{equation} 

$\lambda_\text{pick}\left(t\right)$ is proportional to the rate of \ops\ collisions with the target material or the vacuum cavity:
\begin{equation}
	\lambda_\text{pick}\left(t\right)=n\sigma_\text{a}v\left(t\right)
\end{equation}
where $n$ is the atomic density of the material, $\sigma_\text{a}$ the annihilation cross section, and $v(t)$ the time dependent velocity of \ops. It is thus critical to either eliminate or precisely account for the contribution of pick-off annihilation to the observed decay rate. Another effect that leads to a systematic reduction of the observed lifetime $\lambda_\text{obs}$ is \ops\  escaping the detection region which also depends on its velocity. These are the main limitation of previous experiments. 

The Ann Arbor measurements from Ref.~\cite{Vallery2003} consisted of a positron beam hitting a porous target material known to have a high formation of \ops. Varying the density of the target material and the implantation energy of the positrons to study the behaviour of the pick-off annihilations and \ops\ escaping the detection region, they extrapolated their measurements to find the decay rate in vacuum. However, this introduces the main uncertainty which is of the order of \SI{100}{\ppm}.

The Tokyo group employed a different method~\cite{Asai1995, Jinnouchi2003} which does not require such extrapolation and is therefore free from the above-mentioned systematic error. The setup consists of a $\beta^+$ source surrounded by a silicon powder or gel. The time and energy of the annihilation photons are measured simultaneously with high energy resolution germanium/YAP detectors. The simulated 3$\gamma$ spectrum is subtracted from the data to find the 2$\gamma$ spectrum and determine the time dependence of $\lambda_\text{pick}\left(t\right)/\lambda_{3\gamma}$. After this a fit on the data to determine the vacuum decay rate is performed. The normalization of the 3$\gamma$ distribution introduces the largest uncertainty in the measurement, \SI{90}{\ppm}. Moreover, increasing the source activity would increase pile-up, which severely limits the statistical precision.

Here we present a new method to perform a high precision (\SI{100}{\ppm}) measurement of the decay rate of \ops\  to eliminate the main systematic effects faced in the past by employing the setup developed for our search at ETH Zurich for  \ops\ invisible decay~\cite{Vigo2018}. We use a high intensity positron beam to form \ops\ with a porous SiO$_2$ film and confine it in a vacuum cavity, surrounded by a nearly hermetic, granular calorimeter to detect the annihilation photons. The vacuum cavity is sealed with a thin membrane specifically designed to let positrons through, but confine \ops, therefore suppressing events where \ops\ escapes the detection region. The granularity of the calorimeter allows us to discriminate two- and three-$\gamma$ events, and thus we can obtain the two-$\gamma$ time spectrum due to pick-off annihilations directly from the data. This spectrum is then subtracted from the data prior to the fitting. This method will reduce the systematic effects at a level of \SI{25}{\ppm}. Moreover, the hermeticity of the detector allows us to very effectively veto pile-up events and thus work at high positron rates without the above-mentioned limitations in statistics.



\section{\label{sec:ExpSetup}Experimental Techniques and Setup}

\subsection{\label{subsec:ExpSetup:PsProduction}Positronium Production and Confinement in Vacuum}

\begin{figure}
	\centering
	\includegraphics[width=0.8\linewidth]{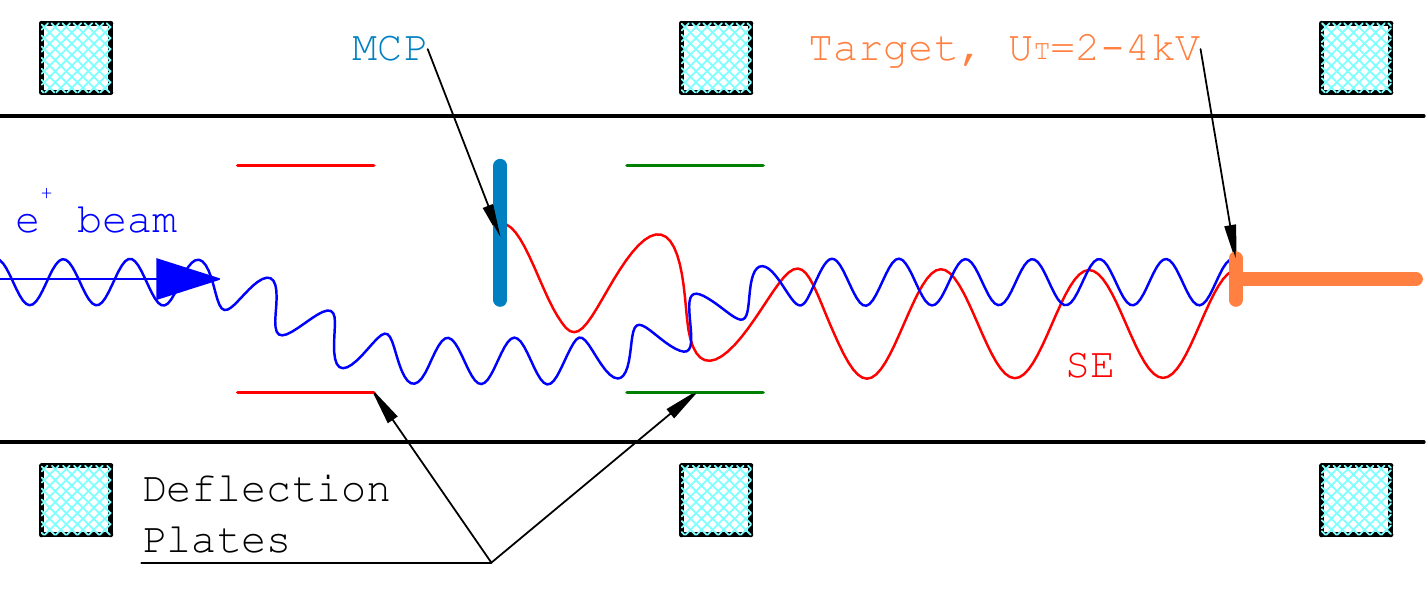}
	\caption{\label{fig:ExpSetup:SketchTagging}Positron tagging scheme with a Micro-Channel Plate~\cite{Vigo2018}. The positron beam (blue helix, coming from the left) is deflected off axis by the deflection plates (red and green) to bypass the micro-channel plate (MCP). Secondary Electrons (SE, red helix) are released when the positron impinges the target and guided back and detected by the MCP. Note that the positron and electron trajectories are only sketches, the actual deflection is perpendicular to the drawing plane.}
\end{figure}

The slow positron beam at ETHZ is based on a \SI{120}{\mega\becquerel} $^{22}$Na radioactive source coupled to a cryogenic Argon moderator, providing a flux of $\Phi_{\si{\positron}}^\text{slow}\sim\SI{4E5}{\positron\per\second}$. The slow positrons are magnetically guided to a SiO$_2$ target, where they are accelerated to \SIrange[range-phrase = --]{2}{5}{\kilo\electronvolt} and either form positronium, i.e.~\ops\ or \pps, or annihilate into $2\gamma$'s. The arrival of a positron to the target is tagged by a Micro-Channel Plate (MCP) which detects Secondary Electrons (SE) released by the impinging positrons, as shown in Fig.\ref{fig:ExpSetup:SketchTagging} (see Ref.~\cite{Alberola2006,VigoPhD} for more details on the beam).

Positronium production into  vacuum is achieved as described in Ref.~\cite{Crivelli2010_2}. Positrons impinge on a porous SiO$_2$ film, \SI{500}{\nano\meter} thick, with a fully interconnected network of pores. The positrons stop in the dense silica matrix, thermalize and form Ps (either \ops\ or \pps), which is expelled into the pores with a kinetic energy of approximately \SI{1}{\electronvolt}, corresponding to the Ps work function in this material~\cite{Nagashima1998} (see Section~\ref{subsec:Errors:ExcitedOps}). \ops\ is then able to diffuse through the pore network and escape through the surface into vacuum, where the Ps energy has a minimum. This process is much faster than the \ops\ lifetime, i.e.~it takes of the order of \SI{1}{\nano\second} for the \ops\ to diffuse into vacuum. However, due to its much shorter lifetime, \pps\ mostly decays within the target and hence in the following we solely consider the \ops.

During the diffusion, the \ops\ suffers of the order of \numrange{E4}{E6} collisions (depending on the implantation depth) with the silica pore walls, losing kinetic energy. However, \ops\ cannot fully thermalize due to quantum mechanical confinement in the pores, and instead the emission energy into vacuum has a minimal value of \SI[separate-uncertainty]{74(5)}{\milli\electronvolt} (C-samples in Ref.~\cite{Crivelli2010_2}), which is reached at positron implantation energies above \SI{3}{\kilo\electronvolt}.

Some \ops\ is emitted without having reached this minimal energy, leading to a high-energy tail extending up to \SI{1}{\electronvolt}, i.e.~the energy at which \ops\ is expelled into the pores. In addition, there is a low-intensity (few \si{\percent}) higher-energy tail up to \SI{100}{\electronvolt} of Ps formed from backscattered positrons that capture an electron at the target surface, which we refer to as fast backscattered positronium~\cite{Vallery2003}.

The collisional wall quenching for low energy of \ops\ is negligible; in fact, \ops\ can clearly survive an order of \num{E6}
collisions while diffusing through the silica film. However, for kinetic energies close to the Ps binding energy of \SI{6.8}{\electronvolt}, the pick-off probability becomes non-negligible, and above that energy the probability can be assumed to be \SI{100}{\percent}. This quenching introduces a non-exponential, 2$\gamma$'s decay rate component $\lambda_\text{pick}\left(t\right)$.

\begin{figure}
	\centering
	\includegraphics[width=0.8\textwidth]{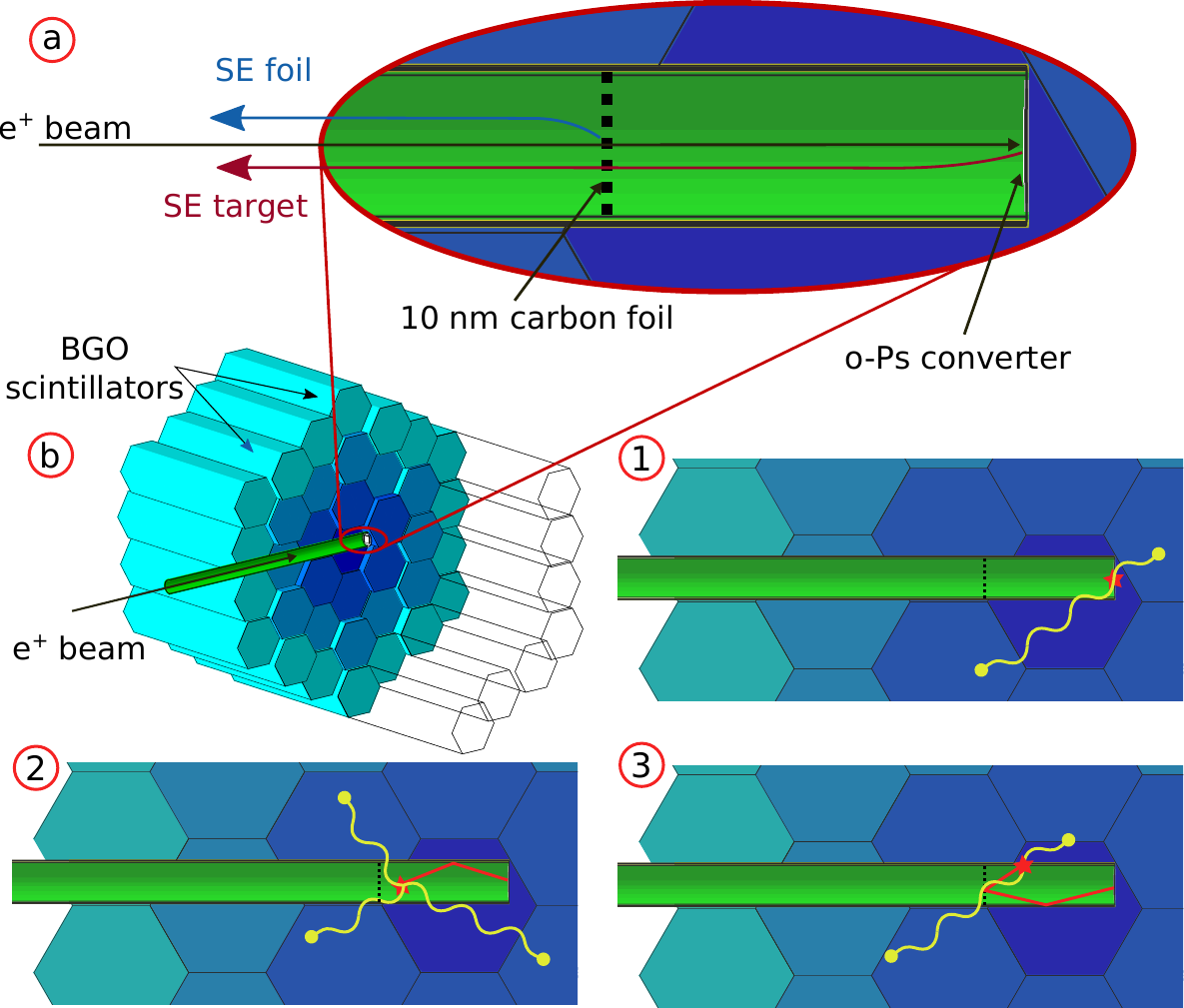}
	\caption{\label{fig:ExpSetup:ECAL}a) Positronium confinement cavity; b) Sectional view of the calorimeter, some scintillators from the cut side are shown as wire-frames for reference. Possible events: $1)$ prompt annihilation and \pps\ decay into two back-to-back \SI{511}{\kilo\electronvolt} photons; $2)$ \ops\ decay into three photons; and $3)$ fast \ops\ quenching after a collision with the cavity walls.}
\end{figure}

\begin{figure}
	\centering
	\includegraphics[width=0.65\textwidth]{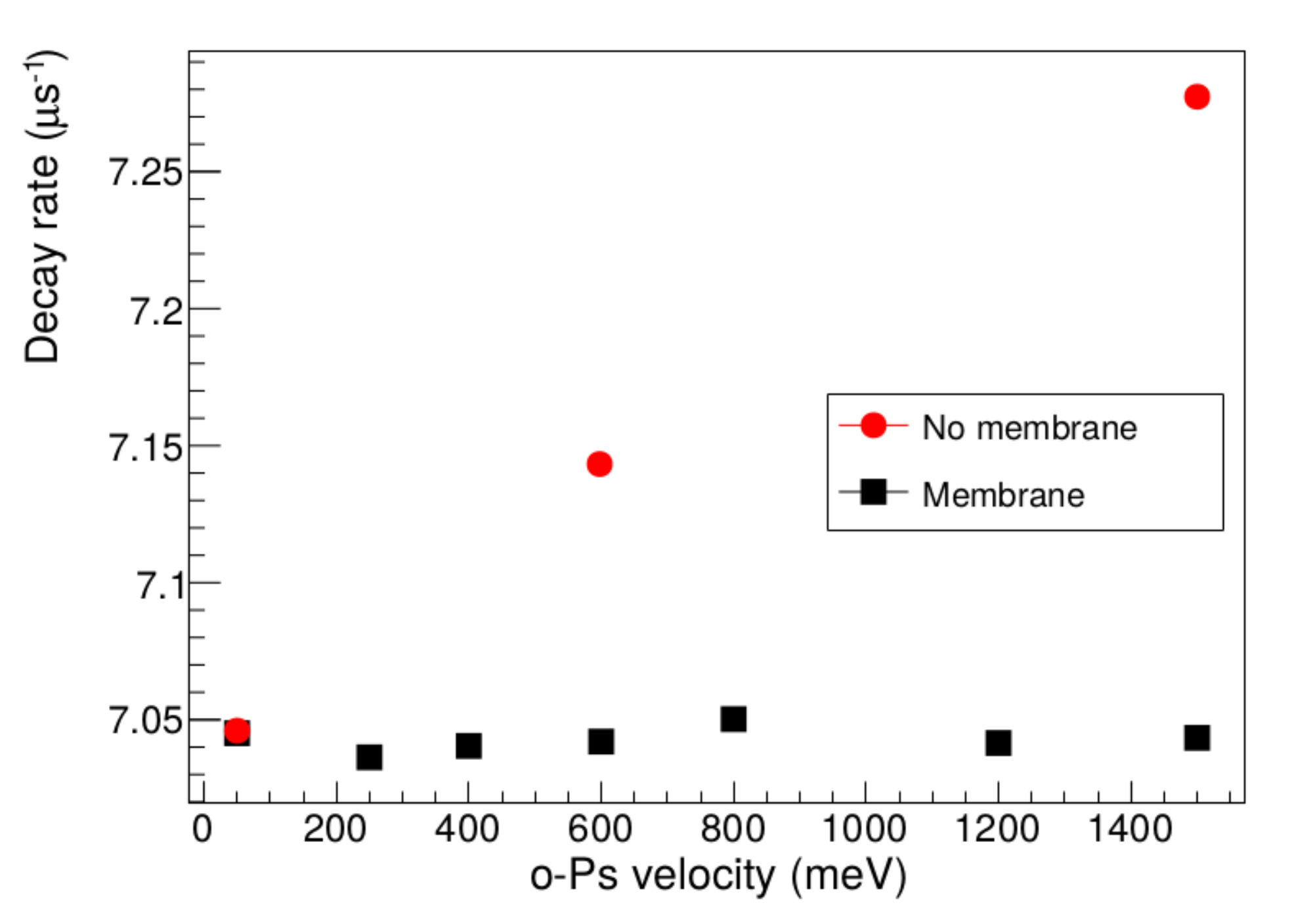}
	\caption{\label{fig:G4Membrane} Simulated decay rate as a function of the \ops\  energy. }
\end{figure}

\begin{figure}
	\centering
	\includegraphics[width=0.65\textwidth]{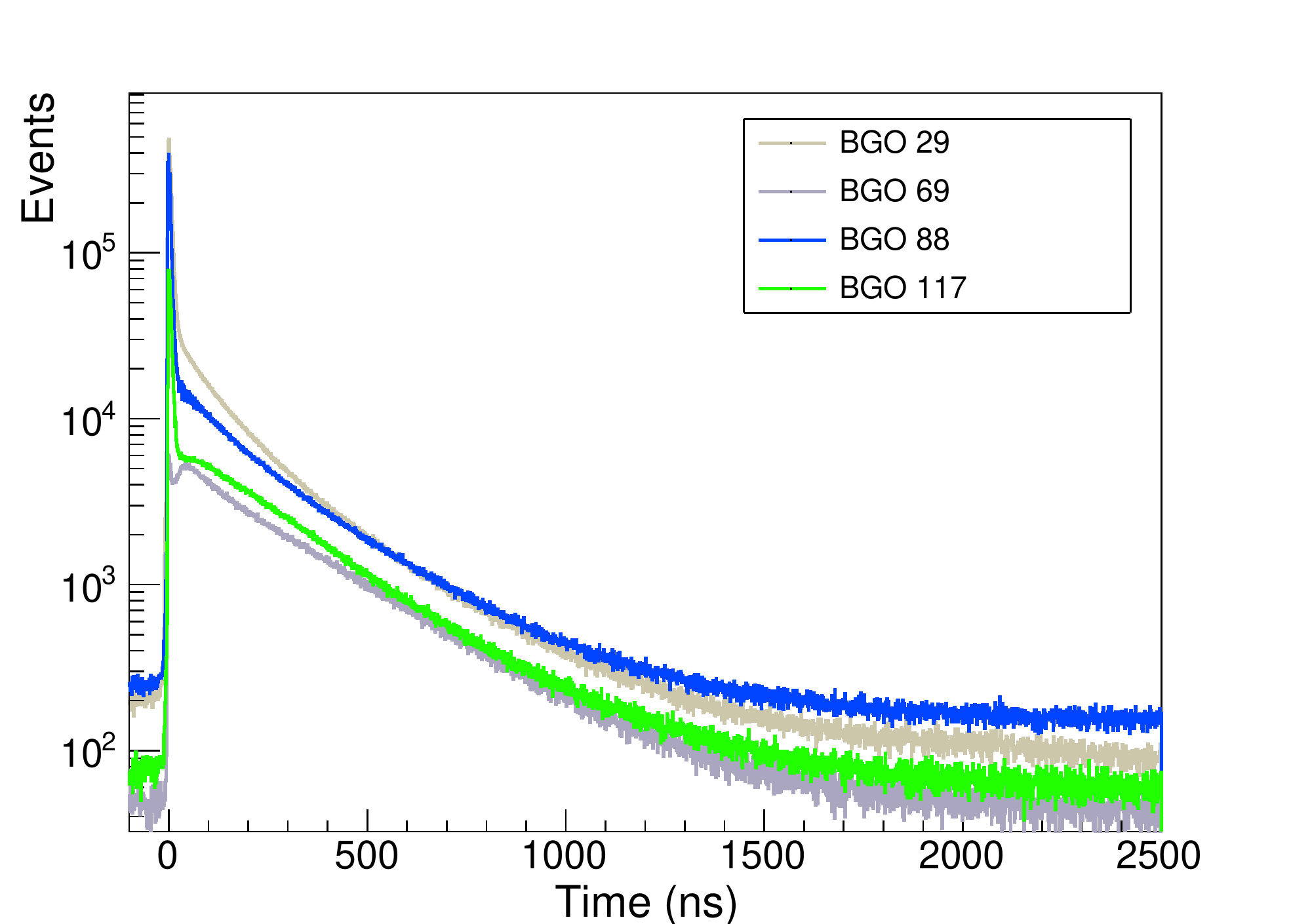}
	\caption{\label{fig:data} Measured time spectra of annihilation photons as a function of the crystal position. Crystal 29: on the target, 69 at 4 cm, 88 at 8 cm and 117 at 12 cm from the target. }
\end{figure}

Fast backscattered \ops\ is also more likely to escape the high detection efficiency region during its lifetime. As explained in Ref.~\cite{Crivelli2010}, we plan to use a \SI{10}{\nano\meter} carbon foil to seal the cavity and prevent \ops\ leakage (see Fig.~\ref{fig:ExpSetup:ECAL}) since this affects the measured lifetime as shown in Fig.~\ref{fig:G4Membrane} and Fig. \ref{fig:data}. Incoming positrons are accelerated to \SI{1}{\kilo\electronvolt} and can thus pass through the thin foil, reaching the target behind. However, even the most energetic \ops\ (up to \SI{100}{\electronvolt}) will be confined within the target, the carbon film and the cavity walls.

Furthermore, with this method an additional signature for the presence of a positron in the formation cavity is added to the trigger scheme. Secondary electrons emitted from both the carbon foil and the target will be detected by the MCP. The positrons are first accelerated to \SI{1}{\kilo\electronvolt} energy by the potential applied between the ground pipe and the carbon foil, thus the SEs emerging from the foil are transported with this energy back to the MCP. In the second acceleration stage, the positrons are accelerated by a \SIrange{1}{3}{\kilo\volt} potential applied between the carbon foil and the target. Once they hit the target, more SEs are emitted and transported back to the carbon foil, where they are energetic enough to pass through and reach the MCP faster than the carbon foil SEs. This technique was already demonstrated in Ref.~\cite{Crivelli2010}.

\subsection{\label{subsec:ExpSetup:ECAL}Annihilation Photons Detection}

We will use the high efficiency, granular calorimeter from Ref.~\cite{Vigo2018,VigoPhD} to measure the time and energy of the annihilation photons (Fig.~\ref{fig:ExpSetup:ECAL}). The detector provides excellent hermeticity and photon detection efficiency, allowing simultaneous fit of time distributions in conjunction with topological cuts, thus, the time dependent pick-off annihilation rate can be measured directly and subtracted from the \ops\ decay rate prior to the fitting of the distribution. This method is free from the systematic error introduced by the extrapolation in the intensity of the backscattered positronium. This subtraction will be \SI{100}{times} smaller than what was applied in the Tokyo measurements and higher statistics can be reached with this method.

\section{Methods and Data Analysis}
\label{sec:Methods}

To reach below \SI{100}{\ppm} statistical precision on the measurement we will need around \num{E9} pure \ops\ events. The \ops\ yield into vacuum is approximately \SI{30}{\percent} at \SI{3}{\kilo\electronvolt}~\cite{Crivelli2010_2} and the expected tagging rate is \SI{2000}{\per\second}~\cite{Crivelli2010}, thus the data acquisition will take in the order of \SI{2}{weeks}. Energy and time information of the annihilation photons will be collected for the 14 darkest blue crystals in Fig.~\ref{fig:ExpSetup:ECAL}, i.e. the two crystals directly next to the target and for the first layer of 12 crystals. The rest of the crystals will only record the deposited energy and will thus be used as a VETO to remove pileup events. 

As described in Section~\ref{sec:ExpSetup}, the largest systematic errors to the \ops\ decay spectrum will be pile-up and pick-off annihilation. Pile-up events can be rejected with an upper cut on the total energy deposited in the calorimeter. 

\begin{figure}
	\centering
	\includegraphics[width=0.7\textwidth]{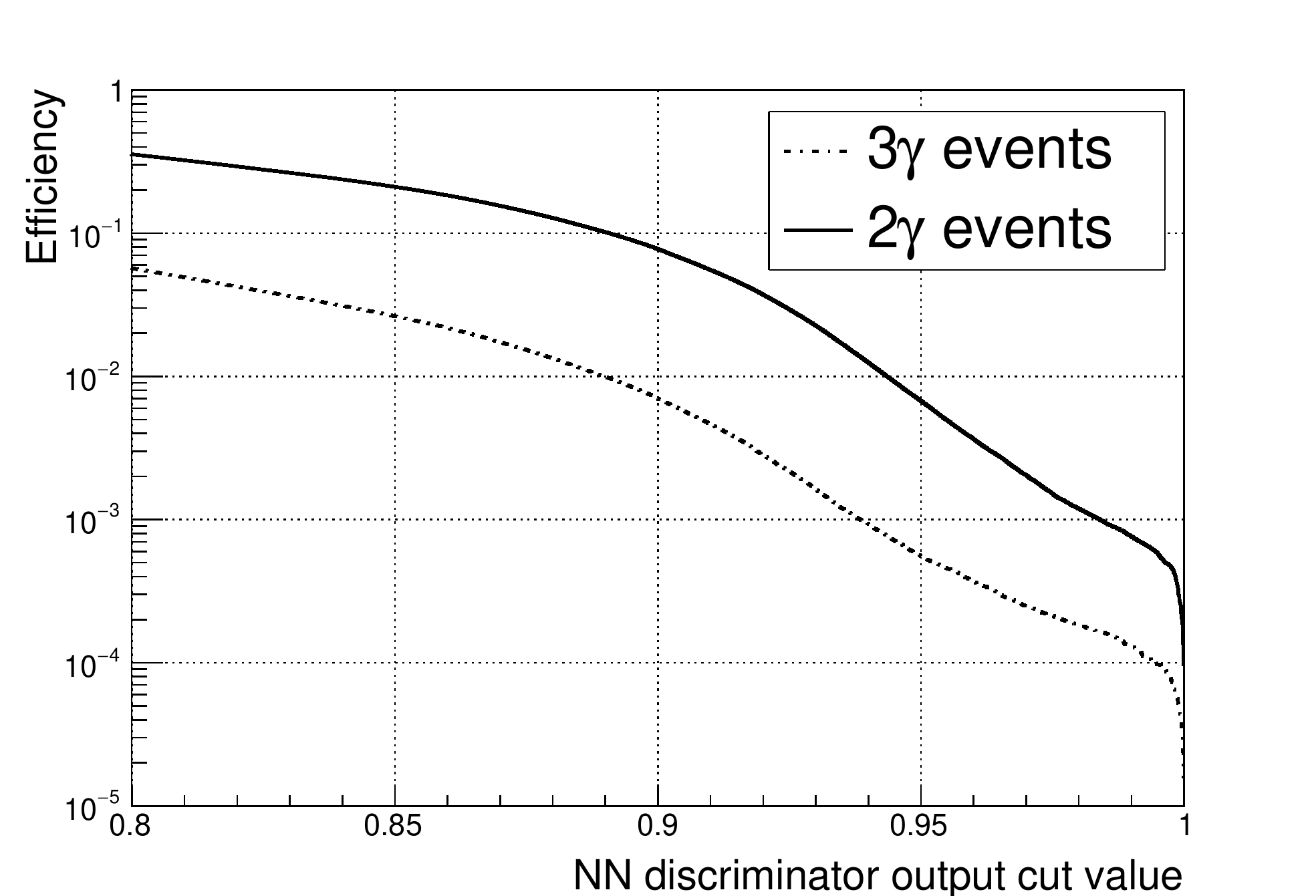}
	\caption{\label{fig:Methods:NN_log}Logarithmic plot of the efficiency for neural network discriminator output value. For a discriminator output around 0.9 the $3\gamma$ can be suppressed below 1\% while maintaining 10\% efficiency for $2\gamma$ events. Data from Monte-Carlo simulations.}
\end{figure}

A neural network has been developed and tested to construct a classifier for $2\gamma$ and $3\gamma$ decay events using data from Monte-Carlo simulations. The $2\gamma$ distribution can be built directly from measurements. The intensity of the backscattered \ops\ ranges from \SIrange{1}{10}{\percent} between \SIrange[range-phrase = ~and~]{2}{5}{\kilo\electronvolt} positron implantation energies~\cite{Vallery2003}. To be able to distinguish between the pick-off and \ops\ time spectra the neural network needs to suppress the $3\gamma$ events below \SI{1}{\percent} efficiency while maintaining above \SI{10}{\percent} efficiency for the $2\gamma$ events. Preliminary training results from the neural network on $2\gamma$ and $3\gamma$ data sets generated with Monte-Carlo simulations are shown in Fig.~\ref{fig:Methods:NN_log}. These results are expected to improve with more data from simulations and measurements, and further training. The neural network can be validated with data by selecting $2\gamma$ and $3\gamma$ samples with timing cuts, i.e.~the prompt peak and \ops\ decay.




The measured $2\gamma$ distribution can then be subtracted from the data, or used directly in the fitting routine. Running the measurement at different implantation energies ranging between \SIrange[range-phrase = ~and~]{2}{5}{\kilo\electronvolt} allows us to verify that the method is independent of the \ops\ velocity, avoiding the requirement to extrapolate to zero backscattered \ops\ intensity.

\section{\label{sec:Errors}Statistical and Systematic Errors}
The total error for the \ops\ decay rate measurement proposed here is expected to be at a level of \SI{104}{\ppm} combined accuracy: $\SI{100}{\ppm}\text{(stat)} + \SI{25}{ppm} \text{(syst)}$. Table \ref{tab:Errors:Summary} summarizes all the expected contributions to the systematic error that were estimated for the experiment, which are described in the next sections. 

\begin{table}
	\sisetup{separate-uncertainty=false}
	\centering
	\caption{\label{tab:Errors:Summary}Summary of expected systematic errors.}
	\begin{tabular}{l r}
		\toprule
		Source of contribution				& Error			\\
		\midrule
		Residuals in the fit				& $<\SI{1}{\ppm}$	\\
		$\gamma$-detection non-uniformity	& $<\SI{1}{\ppm}$	\\
		Subtraction of fast \ops			& \SI{12}{\ppm}		\\
		Trapped \ops						& \SI{9}{\ppm}		\\
		TDC									& $<\SI{15}{\ppm}$	\\
		Time resolution						& \SI{4}{\ppm}		\\
		Uncorrelated \ops\ decays			& \SI{7}{\ppm}		\\
		Ps$^*$								& $<\SI{1}{\ppm}$	\\
		Stark shifting						& \SI{1}{\ppm}		\\
		Magnetic quenching					& \SI{10}{\ppm}		\\
		\midrule
		{\bf Total systematic error}		& \bf{\SI{25}{\ppm}}\\
		\bottomrule
	\end{tabular}
\end{table}

\subsection{\label{subsec:Errors:Stat}Fit of the Decay Curve}
After subtraction of the fast backscattered component and starting to fit at $t>\SI{200}{\nano\second}$ (see below for an explanation of this choice), the time spectrum of the \ops\ is given by:
\begin{equation}
	N(t)=\left(Ae^{-\lambda_\text{T}t} + B\right)e^{-Rt}
	\label{eq:opsspectrum}
\end{equation}
$\lambda_\text{T}$ is the \ops\ decay rate in vacuum, $A$ and $B$ are constants and $R$ is the accidental stop-rate of the detector. This rate, which is due to environmental radioactivity, cosmic background and pile-up, can be measured experimentally. The fit of the spectrum with Eq.~\eqref{eq:opsspectrum} will be performed in the interval \SIrange[range-phrase = --]{200}{1500}{\nano\second}. Hence, to reach \SI{100}{\ppm} (stat.) precision, the number of \ops\ that should be accumulated in the spectrum is of the order of \num{E9}. From simulation, it was found that the systematic error introduced by the fit procedure is less than \SI{1}{\ppm}. 

\subsection{\label{subsec:Errors:DetectionUniformity}Non uniformity of the $\gamma$-detection efficiency}
The non-uniformity of the detection efficiency over the cavity volume clearly affects the exponential slope (e.g.~if younger \ops\ is detected with higher efficiency than the older ones). Simulations show that after a certain time  the \ops\ decays uniformly in the cavity volume. We estimated that for the given cavity after the uniformisation time $t_\text{R}>\SI{180}{\nano\second}$ this effect is negligible ($<\SI{1}{\ppm}$). This is mainly due to the small volume of the cavity that confines the \ops, the optimization of the solid angle covered by the calorimeter along the cavity symmetry axis and the large size of the BGO crystals. Moreover, the \ops\ was simulated with minimal kinetic energy (\SI{75}{\milli\electronvolt}); for higher \ops\ energies, $t_\text{R}$ is even smaller.   

\subsection{\label{subsec:Errors:pick-offSubtraction}Subtraction of the pick-off annihilation rate}
 
\begin{figure}
	\centering
	\includegraphics[width=0.46\textwidth]{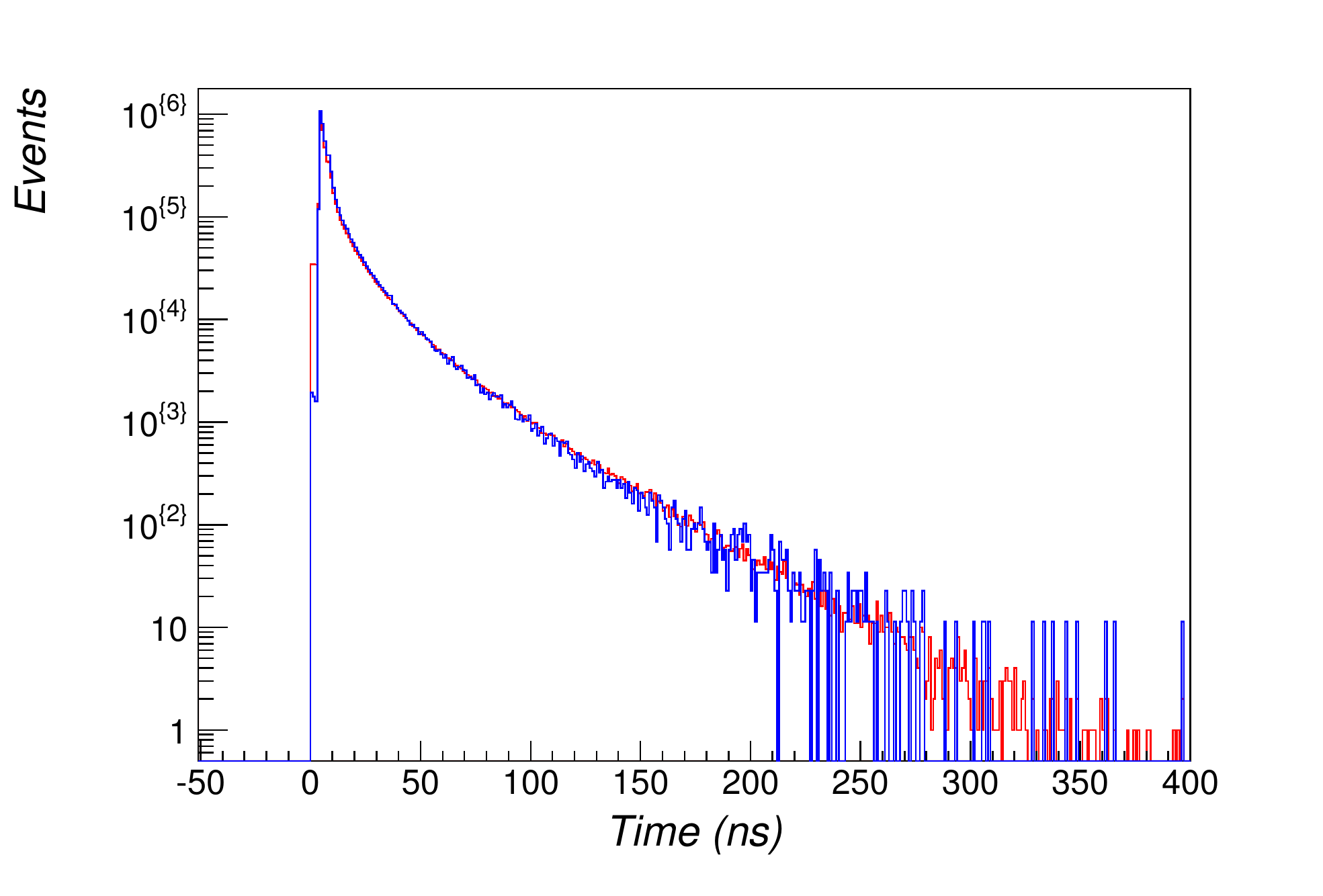}
	\hfill
	\includegraphics[width=0.46\textwidth]{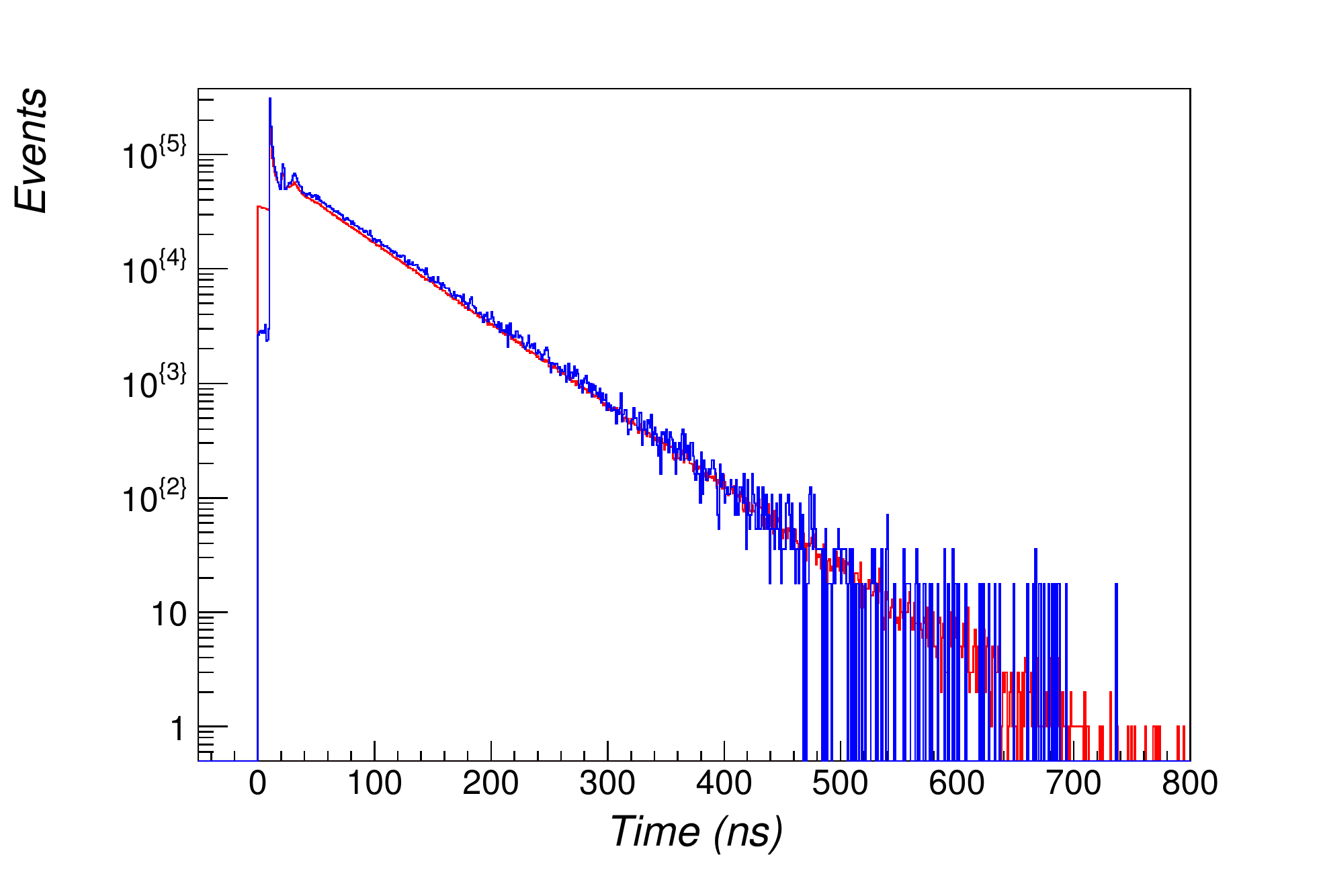}
	\caption{\label{fig:Errors:Backspectra}Time distribution of the collisional annihilation on the cavity walls from the fast backscattered positronium. The two plots have been simulated assuming two different scenarios: (left) linear dependence of the pick-off probability with the energy (until the dissociation energy) and (right) constant probability (0.1) for monoenergetic \ops\ emitted with \SI{3}{\electronvolt}.}
\end{figure}

Fig.~\ref{fig:Errors:Backspectra} shows the simulated time distribution of the collisional annihilation on the cavity walls from the fast backscattered positronium (red line). This spectrum is the sum of the 2 and 3 gamma decays whose fraction depends on the energy distribution of the fast backscattered positronium and on the energy dependence of the pick-off probability. Unfortunately, the knowledge of the last parameter is very poor for the interval between \SI{1}{\electronvolt} (the energy at which \ops\ is formed) and \SI{6.8}{\electronvolt} (the dissociation energy of \ops). We know that it varies from about \num{E-5} to a value close to 1 at the dissociation energy. Therefore, we simulated different scenarios, i.e.~different dependence of the pick-off probability to see how this parameter affects the subtraction method.

\begin{figure}
	\center
	\includegraphics[width=0.46\textwidth]{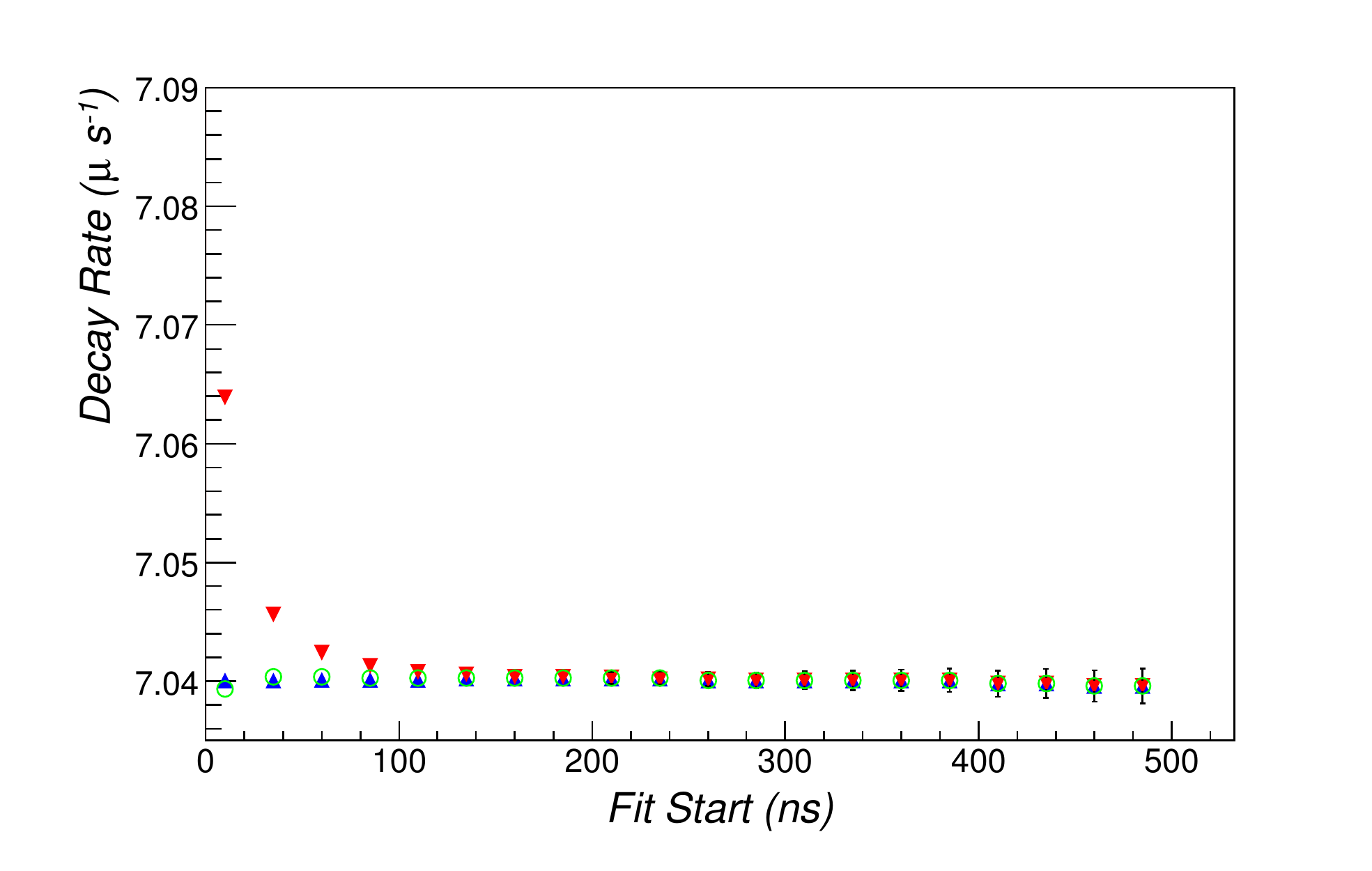}
	\hfill
	\includegraphics[width=0.46\textwidth]{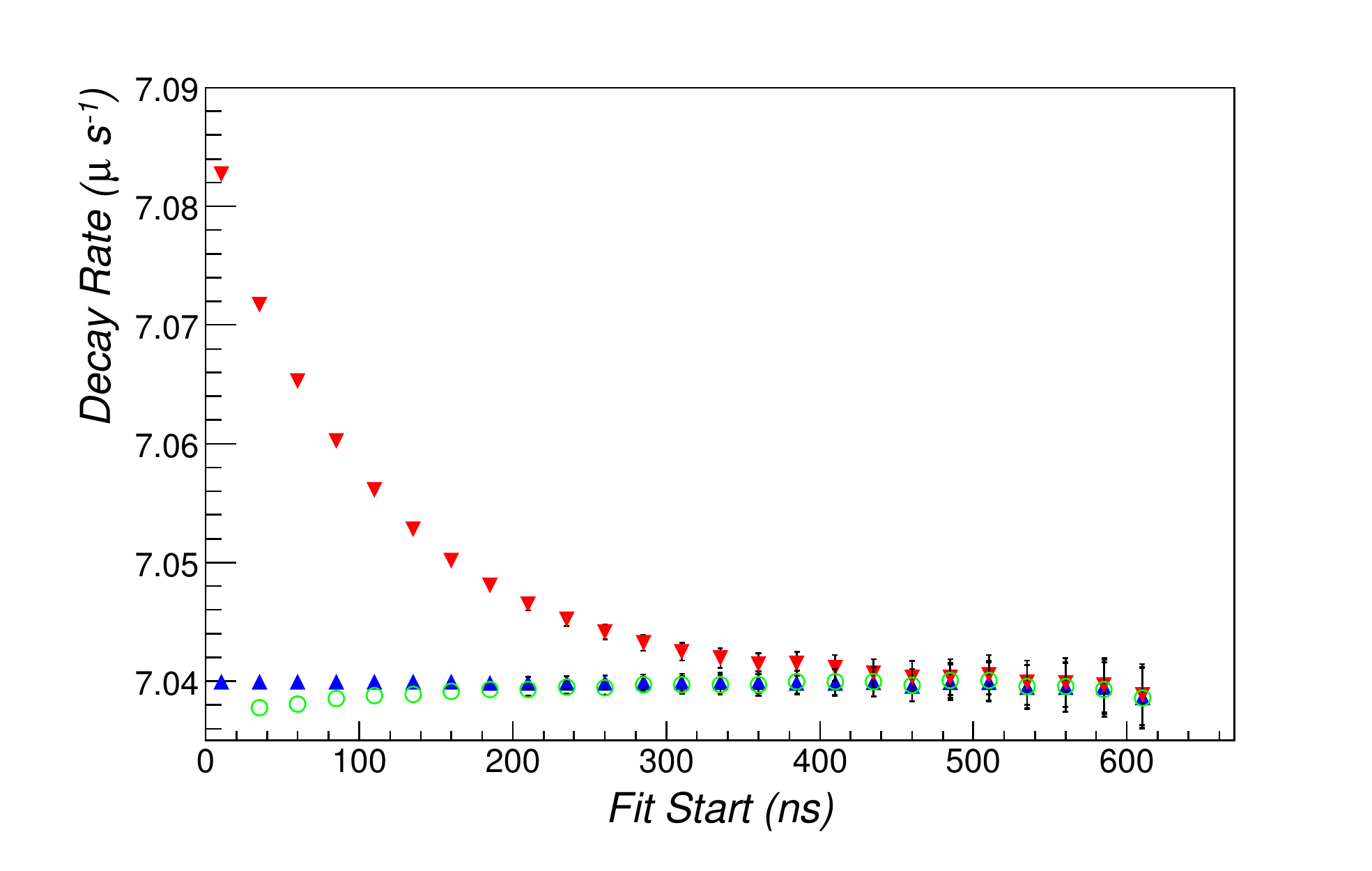}
	\caption{\label{fig:Errors:Subtraction}Decay rate as a function of the fit start for: the sum of the \ops\ in vacuum and the fast backscattered \ops\ (red), the spectrum after the subtraction of this contribution (green) and a pure \ops\ spectrum in vacuum (left and right plots refer to the different scenarios of Fig.\ref{fig:Errors:Backspectra})}
\end{figure}

The blue line on the same plot (Fig.~\ref{fig:Errors:Backspectra}) is the normalized spectra of the back-to-back \SI{511}{\kilo\electronvolt} photons which are detected in the calorimeter. This distribution is subtracted prior to the fitting of the total measured spectrum. The results of this method are presented in Fig.~\ref{fig:Errors:Subtraction}, where the decay rate is plotted as a function of the fitting start time\footnote{From the comparison of these two plots with the one measured by the Michigan group \cite{Vallery2003} one can see that, qualitatively, the simulation reproduces the behavior well.}.

In addition to the back-to back \SI{511}{\kilo\electronvolt} gammas from the fast backscattered \ops, the calorimeter will detect the contributions of the trapped \ops\ and the triplet state quenched by the magnetic field. However, after \SI{200}{\nano\second} (the start of the fit time), their contributions are negligible and, therefore, the fact that their normalization constant will be different will not affect the subtraction.

\begin{figure}
	\centering
	\includegraphics[width=0.46\textwidth]{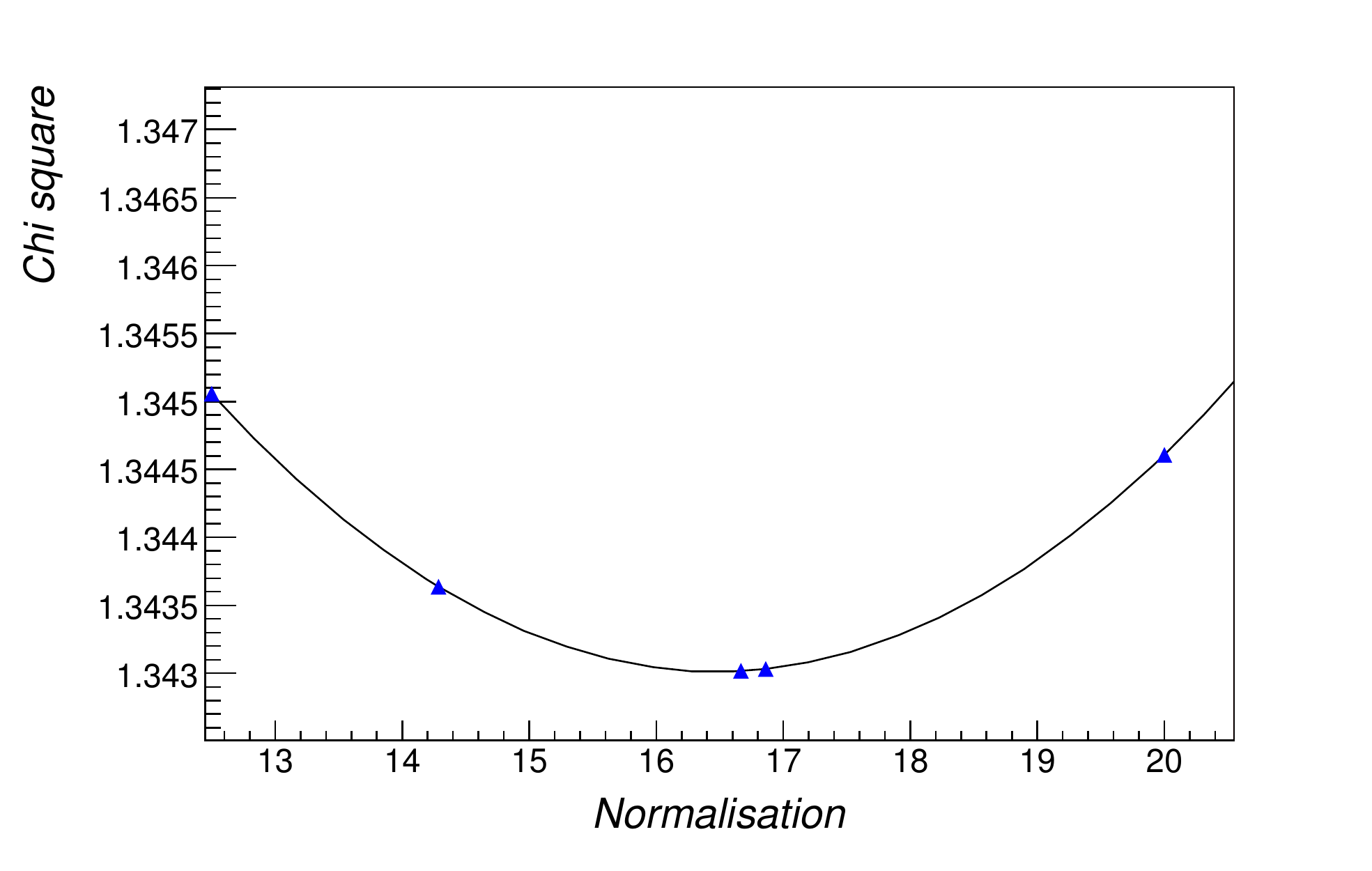}
	\hfill
	\includegraphics[width=0.46\textwidth]{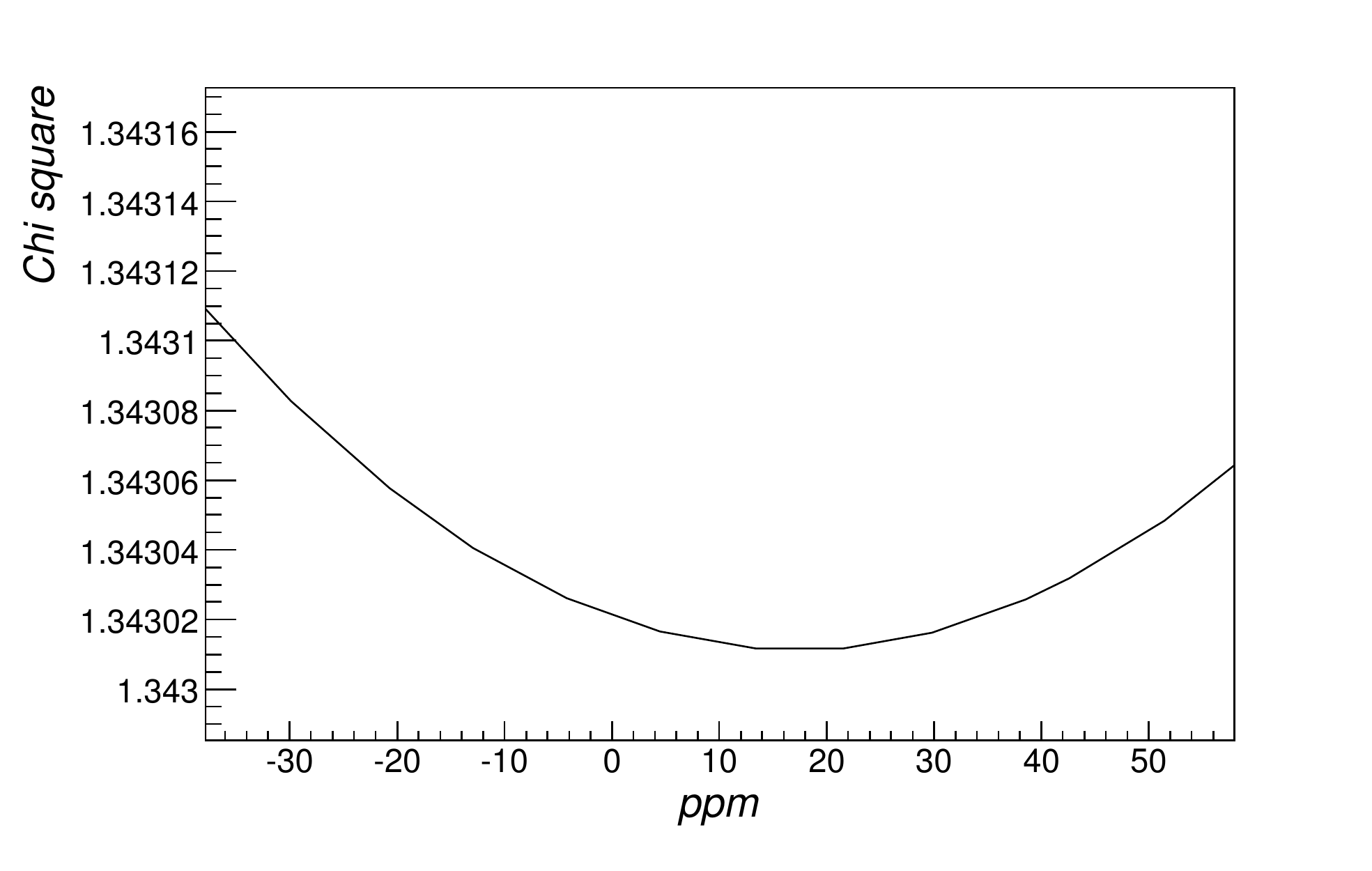}
	\caption{\label{fig:Errors:Chisquare}The $\chi^2$ dependence versus the normalization factor used for the subtraction (left) and the $\chi^2$ versus the systematic error (right).}
\end{figure}

The knowledge of the back-to-back normalization factor $A_\text{B2B}$ clearly affects the systematic uncertainty. To eliminate it one can use the $\chi^2$ of the fit. Figure~\ref{fig:Errors:Chisquare} shows this technique for the most pessimistic scenario in which the fast backscattered \ops\ contribution lasts for a longer time: a) the $\chi^2$ dependence versus normalization factor $A_\text{B2B}\pm\SI{30}{\percent}$ used for the subtraction, b) the $\chi^2$ is plotted versus the systematic error. The systematic error of this technique will be at the level of \SI{12}{\ppm}.

\subsection{\label{subsec:Errors:TrappedOps}\ops\ trapped in the SiO$_2$ film}
As we saw with our measurements, a fraction of the \ops\ produced in our samples is trapped in the SiO$_2$ film where it decays with a shorter lifetime because of the pick-off process. These trapped \ops\ atoms distort the spectrum of the \ops\ annihilation in vacuum. A possible approach is to perform a fitting with a sum of two exponentials. However, for the reasons discussed above the fitting has to start after 200 ns, hence it is preferable to try to minimize this contribution. In fact, the fitting with two exponentials is very delicate; it introduces a systematics and increases the error on the fitted lifetime. Furthermore, the uniformity of the defects and their distribution would play an important role.
  
Therefore, one of the criteria for the selection of the target to be used in the experiment was to have the shortest possible lifetime of the \ops\ trapped in the film (smallest pore size). In our selected sample, this is \SI{12}{\nano\second} with a fraction of \SI{10}{\percent}. Using this values in the simulation and then fitting the spectrum shows that the contribution will be at a level of less than \SI{10}{\ppm}.

\subsection{\label{subsec:Errors:TDC}TDC resolution, calibration and linearity}
The TDC resolution, calibration and the integral non-linearity contribute as well to the systematics. It was shown\cite{Jinnouchi2003} that with modern TDCs, this effect is below \SI{15}{\ppm}.

\subsection{\label{subsec:Errors:Tagging}Time resolution of the tagging system and of the BGO crystals}
The influence of the tagging-BGO time resolution affects very weakly the slope of the decay curve. For the expected time resolution of \SI{3}{\nano\second}, the affection of the measured slope will be at \SI{6}{\ppm}.

\subsection{\label{subsec:Errors:UncorrelatedDecay}Uncorrelated \ops\ decay}
In case of positron pile-up, the resulting spectrum will have an additional exponential component decaying with twice the \ops\ decay rate: 
\begin{equation}
	N(t)=(Ae^{-\lambda t}+B+Ce^{-2\lambda t})e^{-Rt}
	\label{eq:uncorrelatedoPs}
\end{equation}
The rate at which two \ops\ are in the cavity at the same time is given by:
\begin{equation}
	R_{2\ops}=2\cdot R_{\si{\positron}} \cdot T_\text{G} \cdot \epsilon_{\ops} 
	\label{eq:uncorrelatedoPsProb}
\end{equation}
where  $R_{\si{\positron}}=\SI{4E5}{\positron\per\second}$ is the positron flux, $T_\text{G}=2~\mu$s is the gate needed to perform the measurement and  $\epsilon_{\ops}$ the efficiency of \ops\ production. Simulations show that fitting the measured spectra with Eq.~\eqref{eq:uncorrelatedoPs} is not a good approach because in this case the systematics will be at \SI{98}{\ppm} and the error on the fitted parameter increases approximately by a factor 2.
  
To avoid this problem the Tokyo group used a source with very low activity (see Section~\ref{sec:Intro}); however, to increase the statistics a beam based experiment is preferable. In fact, with a beam one can select only the annihilations from positrons with a valid start signal in the detector.

The coincidence from the SE in the MCP will be used to generate a signal of the pulse width necessary to perform the measurement that blocks the possibility of other positrons to enter the cavity during this time. In our setup, since the SEs need about \SI{20}{\nano\second} to reach the MCP, there will be a gap before the gating to be effective. Moreover, an upper cut on the total energy measured by the calorimeter will veto very effectively any pile-up event. From simulations the systematic error introduced by this effect is estimated to be below \SI{1}{\ppm}.

\subsection{\label{subsec:Errors:ExcitedOps}Positronium in excited states}

The work function for positronium in SiO$_2$ is

\begin{equation}
W_\text{Ps} (n)= \mu_\text{Ps} + E_\text{B} - \frac{E_\text{Ps}}{n^2}.
\end{equation}
where $\mu_\text{Ps}$ is the chemical potential of Ps in SiO$_2$, $E_\text{Ps}$ its binding energy in SiO$_2$ and $E_\text{Ps}$ its binding energy in vacuum. Any state other than the ground state that has $W_\text{Ps} (n=1)=  \SI{-1}{\electronvolt}$ will thus have a positive work function and will not be emitted from the bulk into the pores.

However, similar to fast backscattered \ops, Ps in an excited state could be formed at the pore surface. For $n=2$ levels, the three P-states decay to the ground state with a radiative lifetime of \SI{3.2}{\nano\second}, and the 2$^1$S$_0$ state has an annihilation lifetime of \SI{1.0}{\nano\second}. The lifetime of these states is too short to contribute to the triplet ground state. 

The most dangerous contribution could arise from the 2$^3$S$_1$ state due to its natural annihilation lifetime of \SI{1.1}{\micro\second} and a formation probability in SiO$_2$ of \num{3E-4}~\cite{Hatamian1987}. However, after its formation the 2$^3$S$_0$ state undergoes about \numrange{E4}{E6} collisions in diffusing through the silica film before it escapes through the surface into vacuum. During this process, the probability that it collisionally quenches to the 2P states (which are separated by only \SI{E-5}{\electronvolt}) is close to 1. Furthermore, the electric field applied to the cavity (\SIrange[range-phrase = --]{3}{4}{\kilo\volt}) quenches the triplet states via Stark mixing with the 2P state. The probability for fast backscattered positronium in 2$^3$S$_0$ to dissociate after the first collision can be assumed to be \SI{100}{\percent} because its binding energy (\SI{1.7}{\electronvolt}) is close to the minimum of the energy (\SI{2}{\electronvolt}) at which the fast \ops\ is emitted.  Therefore, the effect of the  2$^3$S$_0$ is considered to be negligible.

A contribution from higher excited states is even smaller. In fact, the production rate is suppressed by approx.~$\frac{1}{n^3}$, hence the intensity for $n>2$ states is further reduced. Additionally, collisional and dissociation quenchings are even stronger because these highly excited Ps are more fragile. The conclusion is that Ps$^*$ will not affect our decay rate measurement of the triplet ground state. 

\subsection{\label{subsec:Errors:Stark}Stark shift quenching}
The Stark shift stretches the lifetime of Ps atoms, i.e.~a perturbative calculation \cite{Sewell1949} shows that the shift is proportional to the square of the effective electric field $E$ such that
\begin{equation}
\frac{\Delta\lambda_{3\gamma}}{\lambda_{3\gamma}} = 248\left(\frac{E}{E_0}\right)^2
\end{equation}
where $E_0 = \frac{m^2_ee^5}{\hbar^4}\simeq\SI{5.14E9}{\volt\per\centi\meter}$. $E$ is defined as the root-mean-square electric field sensed by \ops\ during its lifetime. In our experiment, this will be about \SIrange[range-phrase = --]{3}{4}{\kilo\volt\per\centi\meter}, and therefore this contribution will be at a negligible level (\SI{1}{\ppb}).

\subsection{\label{subsec:Errors:Magnetic}Positronium quenching in magnetic fields}
Another quenching of \ops\ to \pps\ (and vice versa) is caused by the presence of a magnetic field. The triplet state with $S_\text{z}=0$ can mix with the singlet state resulting in a reduction of the observed lifetime of \ops. The triplet states of \ops\ with  $S_\text{z}=1$ are not affected by the magnetic field; therefore, the maximum reduction in the \ops\ fraction is $\frac{1}{3}$.

For a constant magnetic field the decay rates for the perturbed states are~\cite{Halpern1954}:
\begin{equation}
	\begin{split}  
		\Gamma_{t}(B_z)=\Gamma_{\ops\to 3\gamma}+\Gamma'\left(\frac{\mu_\text{B} B_z}{\Delta E_\text{hf}}\right)^2 \\
		\Gamma_{s}(B_z)=\Gamma_{\pps\to 2\gamma}-\Gamma'\left(\frac{\mu_\text{B} B_z}{\Delta E_\text{hf}}\right)^2
	\end{split}
\end{equation}
where $\Gamma'=\Gamma_{2\gamma}-\Gamma_{3\gamma}$,  $B_z$ is the field strength, $\mu_\text{B} = \frac{eh}{2\pi mc} = \SI{9.27E-24}{\angstrom^2}$ is the Bohr magneton and $\Delta E_\text{hf}= \SI{8.4E-4}{\electronvolt}$ is the energy splitting of the unperturbed ground state. At the field of \SI{20}{\gauss} expected in the cavity the lifetime of \ops\ in the $S_z=0$ state is shifted downwards by \SI{21}{\ppm}.


The fitting of the simulated distributions, which were performed considering that only the triplet state with $S_z=0$ can mix with the
singlet state, predict a systematic shift of -10 ppm on the measured decay rate. The 2 photons annihilations from this quenching are detected in the calorimeter and will be subtracted together with the fast backscattered \ops. Due to their low fraction this will not affect the statistics.   
\section{\label{sec:Summary}Summary}

We proposed a precise determination of the ortho-positronium (\ops) decay rate in the context of the search for dark matter ongoing at ETH Zurich via \ops\ invisible decay. The decay rate experiment aims for an accuracy of \SI{100}{\ppm} to be able to test the second-order correction in the calculations, which is $\simeq 45\left(\frac{\alpha}{\pi}\right)^2\approx\SI{200}{\ppm}$. The precision will be limited by the statistics, thus, if the expectations are fulfilled, this experiment could pave the way to reach the ultimate accuracy of a few \si{ppm} level to confirm or confront directly the higher order QED correction.

\section*{\label{sec:Acknowledgements}Acknowledgements}
This work was supported by ETH Zurich under the grant ETH--35--14--2. We gratefully acknowledge D.~Gidley, R.~Vallery, S.~Gninenko and S.~Belov for very useful and interesting discussions. We are grateful to D.~Cooke for his essential developments in the early stage of this work. We thank L.~Laszlo for providing the thin silica films used for positronium production.

\section*{\label{sec:References}References}
\bibliographystyle{iopart-num}
\bibliography{bibliography} 

\end{document}